# Fracture Mechanics and Self-Healing in Molecular Crystal Resonators for Lasing

Melchi Chosenyah,[1] Vuppu Vinay Pradeep,[1] Kapil Yadav,[1] Ankur Khapre,[1] Biswajit Kumar Barman,[1] Raghavaiah Pallepogu,[2] Rajadurai Chandrasekar[1,✉] (ORCID: 0000-0003-3527-3984)

[1]Advanced Photonic Materials and Technology Laboratory, School of Chemistry and Centre for Nanotechnology, University of Hyderabad, Gachibowli, Hyderabad 500046, Telangana, India.

E-mail: r.chandrasekar@uohyd.ac.in

[2]Department of Chemistry, School of Chemical Sciences, Central University of Karnataka, Kadaganchi, Aland Road, Kalaburagi - 585367, Karnataka, India.

## Abstract

Microscale damage poses a significant challenge to the reliability and longevity of miniaturized photonic devices; however, the mechanisms and repair of microscopic cracks and fractures in organic optical crystals remain largely unexplored. Here, we employ a mechanophotonic approach to induce microscale cracks and complete fractures in single crystals of 3-[4-(dimethylamino)phenyl]-1-(2-hydroxyphenyl)prop-2-en-1-one (HDMAC) using an atomic force microscopy (AFM) cantilever tip. We demonstrate the self-healing capability of these crystals while retaining their optical waveguiding and resonator properties, which are essential for lasing applications. HDMAC microcrystals exhibit self-healing of cracks introduced along the (010) and (001) crystallographic planes, demonstrating the versatility of this phenomenon across distinct crystal orientations. Furthermore, we identify different fracture mechanisms and establish a comprehensive assessment of healing efficiency by integrating optical loss measurements, resonator characteristics, and nanomechanical analyses. This multifaceted approach enables fair evaluation of the recovery of structural integrity and photonic functionality following self-healing.

## Introduction

The ability of crystals to self-repair cracks and restore their structural integrity following mechanical damage or applied stress is a rare and remarkable phenomenon. Notable classical examples of crack healing have previously been reported in polycrystalline $Al_2O_3$ and single crystals of rock salt and KCl.[1,2,3] The self-healing phenomenon has been observed in several molecular crystals[4,5,6,7,8,9,10,11,12,13,14,15] and is linked to dynamic covalent chemistry, diffusion, phase changes, intercalation, electrostatic interactions, and photochemical reactions, highlighting the diverse pathways through which crystal healing can occur. The technological applicability of self-healing crystals depends on both intrinsic factors, such as the healing mechanism, photonic and optoelectronic attributes, and extrinsic factors, including crystal shape and size, nature of crack, crack scale, healing efficiency, and repair rate. Further, if the crystal healing occurs after several hours of contact and requires constant application of compressive stress, its practical applicability is limited.

Single crystals of dipyrazolethiuram disulfide, about 3 mm in length, exhibited a meager ~6.7% healing efficiency after being subjected to a mechanical compressive force of 1-3 N for 24 h, with visible crack marks remaining even after the healing process.[4] The reversible boronic ester reactions enable millimeter-sized crystals to achieve a healing efficiency of up to 67% under compressive stress.[6] The millimeter-long, needle-shaped, 3,3’,5,5’-tetramethyl-4,4’-bipyrazole[8] and dimethyl-4,4'-(methylenebis(azanediyl))dibenzoate[9] crystals showed autonomous healing of cracks and healing via rapid recombination of crystal fragments in just milliseconds with crystallographic precision.

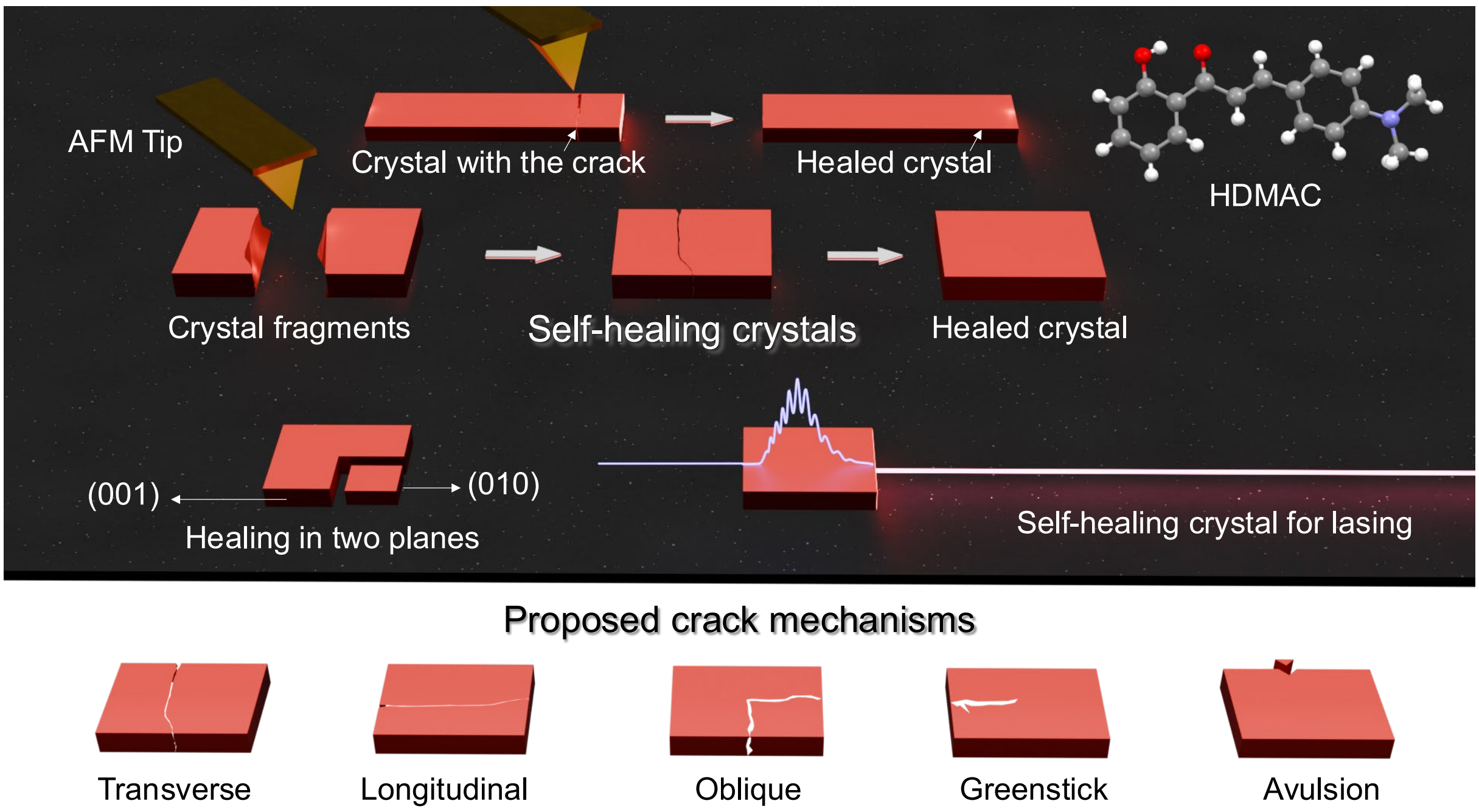

**Fig. 1 | Schematic of mechanophotonics-assisted self-healing in HDMAC microcrystals and proposed fracture mechanics in crystals.** Schematic illustration of micromechanically (mechanophotonics) assisted self-healing of HDMAC molecular crystals across the (001) and (010) planes, with restoration of photonic functionality and proposed crack mechanisms in the crystals.

The healing efficiency and healing time of the crystal are dependent on the type and morphology of the crack, as different fracture modes produce different fracture-surface characteristics. By analogy with the bone fracture, the crystal crack can be broadly classified as transverse (crack across the crystal growth axis with/without crystal's complete separation), longitudinal (crack along the crystal growth axis with/without crystal's complete separation), oblique (diagonal, with/without crystal's complete separation), avulsion (involving the detachment of a small fragment or a piece or material loss), and greenstick (the crystal bends or deforms along the crack without complete separation).[16-18]

To date, self-healing has predominantly been demonstrated in millimeter- to centimeter-sized crystals using the forceps-and-needle technique, typically across a particular crystal facet.[4,6,8,9] In contrast, submicron- and nano-scale crystals with their smooth, mirror-like light-reflecting facets,[19,20,21] can trap light and guide light,[22,23,24] enabling them to function as photonic resonators for lasing and other optical applications. Therefore, investigating self-healing down to these length scales, across multiple crystal facets, is critical for advancing their photonic device functionality.[25,26] However, such studies remain challenging, owing to the need for precise mechanical micromanipulation and controlled contact of tiny crystals. For example, microcrystals of 3-[4-(dimethylamino)phenyl]-1-(2-hydroxyphenyl)prop-2-en-1-one (HDMAC) display lasing characteristics.[27] Our current study reveals that the mechanically cracked HDMAC microcrystals can undergo self-healing while recovering their photonic performance.

In this study, we demonstrate micromechanically induced cracking and subsequent self-healing in HDMAC single-crystal optical waveguides and resonators down to submicron and nanoscale dimensions. Employing a mechanophotonics approach,[28,29] an atomic force microscopy (AFM) cantilever tip is used to precisely generate various types of crystal cracks and to apply controlled compressive force to the cracked crystal, which promotes crack closure and healing. HDMAC exhibits remarkable self-healing capability for mechanically induced microscopic cracks oriented approximately perpendicular to both the (001) and (010) crystallographic planes, indicating that the healing process is accessible through different crystal orientations. Optical excitation of microcrystals displayed self-guiding orange-red fluorescence (FL) along the crystal axis, and also crystal-size-dependent optical resonant modes in their FL spectra.[27] Remarkably, mechanical cracking of these crystals results in the disappearance of their resonant modes, which are restored upon self-healing, highlighting HDMAC microcrystals' potential for mechanically resilient self-healing micro lasers (Fig. 1).

**Results**

The HDMAC crystallizes in ethanol to form thin, millimeter-sized rectangular crystals.[27] When exposed to UV light, the crystals exhibited an orange-red FL with the spectrum centered at 660 nm (Fig. 2a,b). The single-crystal X-ray structure of the HDMAC crystal showed the monoclinic space group P21/c (Supplementary Table 1). In the crystal lattice, the molecules exhibit intramolecular O-H···O hydrogen bonding (1.779 Å) (Fig. 2c) and two different types of intermolecular C-H···O hydrogen bonding (2.666 Å, 2.763Å) (Fig. 2f). The intermolecular interactions are further visualized by Hirshfeld surface analysis using CrystalExplorer21 software, where bright red spots on the $d_{norm}$ surface indicate strong O-H···O intermolecular hydrogen-bonding interactions (Supplementary Fig. 1). The molecule demonstrated nearly planar geometry with minimal torsion angle between its double bond and the benzene rings (3.52° and -7.38°) (Fig. 2c). The crystallographic analysis of HDMAC revealed a step-like molecular orientation along the *b*-axis (Fig. 2e). The face indexing was done using the Crysalispro software, and the facets were found to be (100), (010), and (001), which matched the reported literature[27] (Fig. 2d-f and Supplementary Fig. 2).

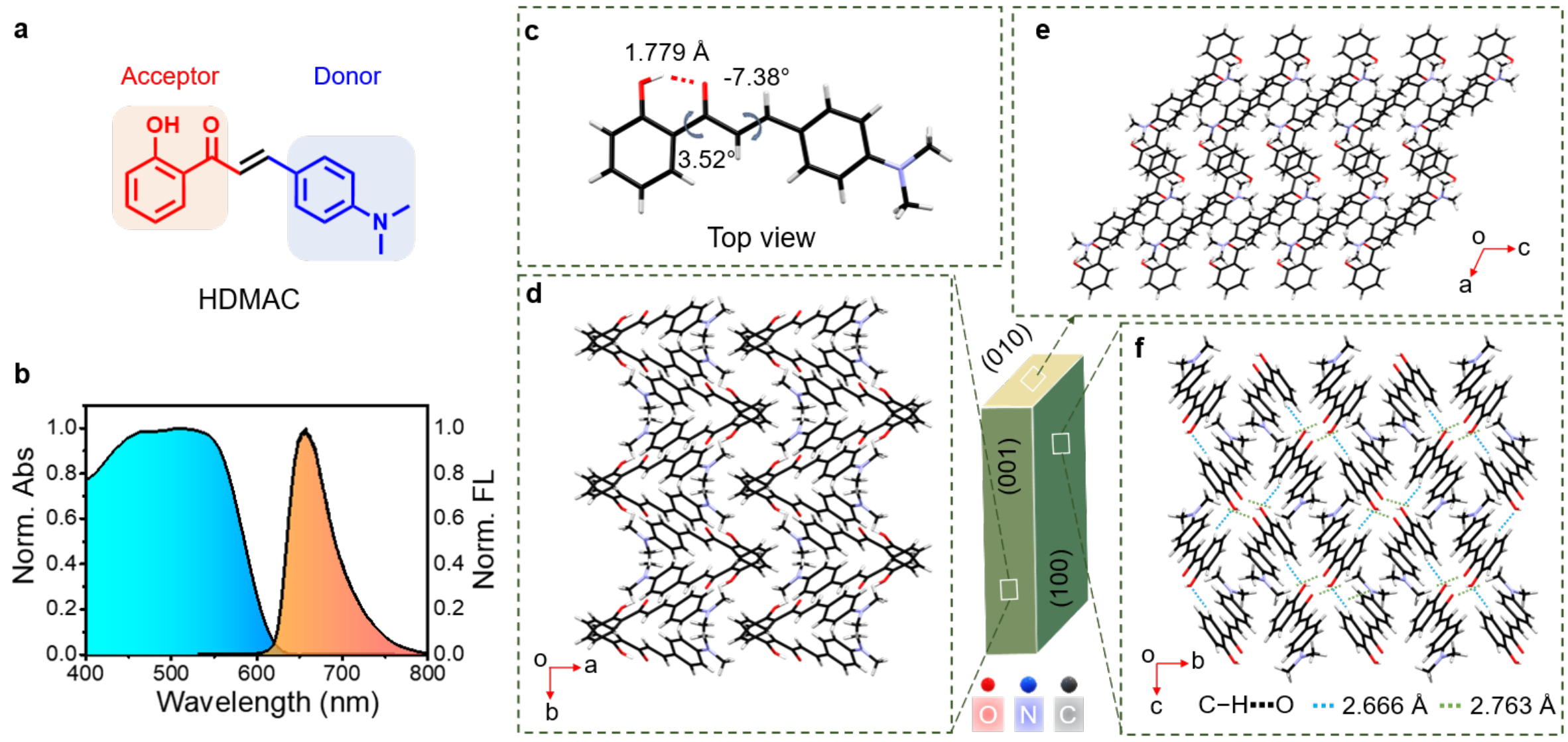

**Fig. 2 │ Molecular structure, solid-state photophysical properties, and single-crystal X-ray diffraction analysis of HDMAC.** **a** Molecular structure of HDMAC. **b** Normalized absorption and emission spectra of HDMAC in the solid state. **c** Intramolecular hydrogen-bonding interactions and torsional angles of the HDMAC molecule. **d-f** The crystal packing of HDMAC along (001), (010), and (100) facets corresponds to the crystallographic *c*, *b*, and *a* axes, respectively. The inset shows a graphical illustration of the crystal with its facets.[27]

The microcrystals were grown via self-assembly from a 2 mg/mL solution of HDMAC in ethanol for mechanical and photonic studies. The solution was drop-cast onto a clean glass coverslip, and the solvent was allowed to evaporate under ambient conditions. The optical microscopy images revealed 2D square and rectangular microcrystals (Supplementary Fig. 3). Under the confocal microscope (Supplementary Fig. 4), using an AFM cantilever tip, an irregular, diagonally propagating microscopic oblique crack was introduced into an HDMAC microcrystal A with dimensions of ≈13.7×12.5 $\mu m^2$ by applying a gentle force along [010]. The crystal exhibited fracture with a combination of oblique and avulsion mechanisms. Subsequently, when the side of the crystal was pushed inward with the cantilever tip, the oblique crack disappeared, indicating the crystal's self-healing nature (Fig. 3a, b). The FESEM micrograph revealed a nearly seamless repair of the oblique crack (inset, Fig. 3c), with only a faint residual line discernible in the healed region at higher magnification. A localized unhealed area persists at the site of avulsion, due to a missing crystal fragment (inset, Fig. 3c). To verify the integrity of the healed interface, another microcrystal S1 was mechanically pushed with the AFM cantilever tip after healing. The entire microcrystal moved as a single body without separation at the healed interface, demonstrating that the crack had been effectively healed and that robust mechanical continuity had been restored (Supplementary Fig. 5 and Supplementary video 1). To examine whether the same self-healing behavior is exhibited by macrocrystals, a thin flake-like macrocrystal was selected for the experiments. Owing to its flake-like structure, the crystal shattered upon the application of mechanical force instead of developing a well-defined crack. Consequently, the fractured interface contained substantial avulsion and irregular fracture surfaces, preventing precise realignment between the broken parts. As a result, self-healing was not observed in macrocrystals (Supplementary Fig. 6), despite possessing the same crystallographic packing as the microcrystal Supplementary Fig. 7).

To investigate the self-healing of two broken pieces, a rectangular-shaped microcrystal B with dimensions of ≈19.7×11.5 $\mu m^2$ was chosen (Fig. 3e). A relatively higher force was applied to the center of the (001) facet using the cantilever tip, deliberately splitting the crystal into two separate fragments (1 and 2) with the exposed irregular broken (010) surfaces. It created a transverse crack and an avulsion at the tip-contact area (Fig. 3d,e). To examine the self-healing of the exposed broken regions, fragment 2 was carefully positioned to align its irregularly broken surface with that of fragment 1. A small additional force applied to the (010) sides of crystal fragment 1 brought the fragments into contact, resulting in immediate healing (Fig. 3d,e, and Supplementary video 2). The FESEM images established the self-healing of the fragments and the avulsion area (inset, Fig. 3f). However, a close-up view revealed a lattice mismatch of ≈250 nm at the edge of the healed microcrystal, which was observed to result from a crystal misalignment (Fig. 3f).

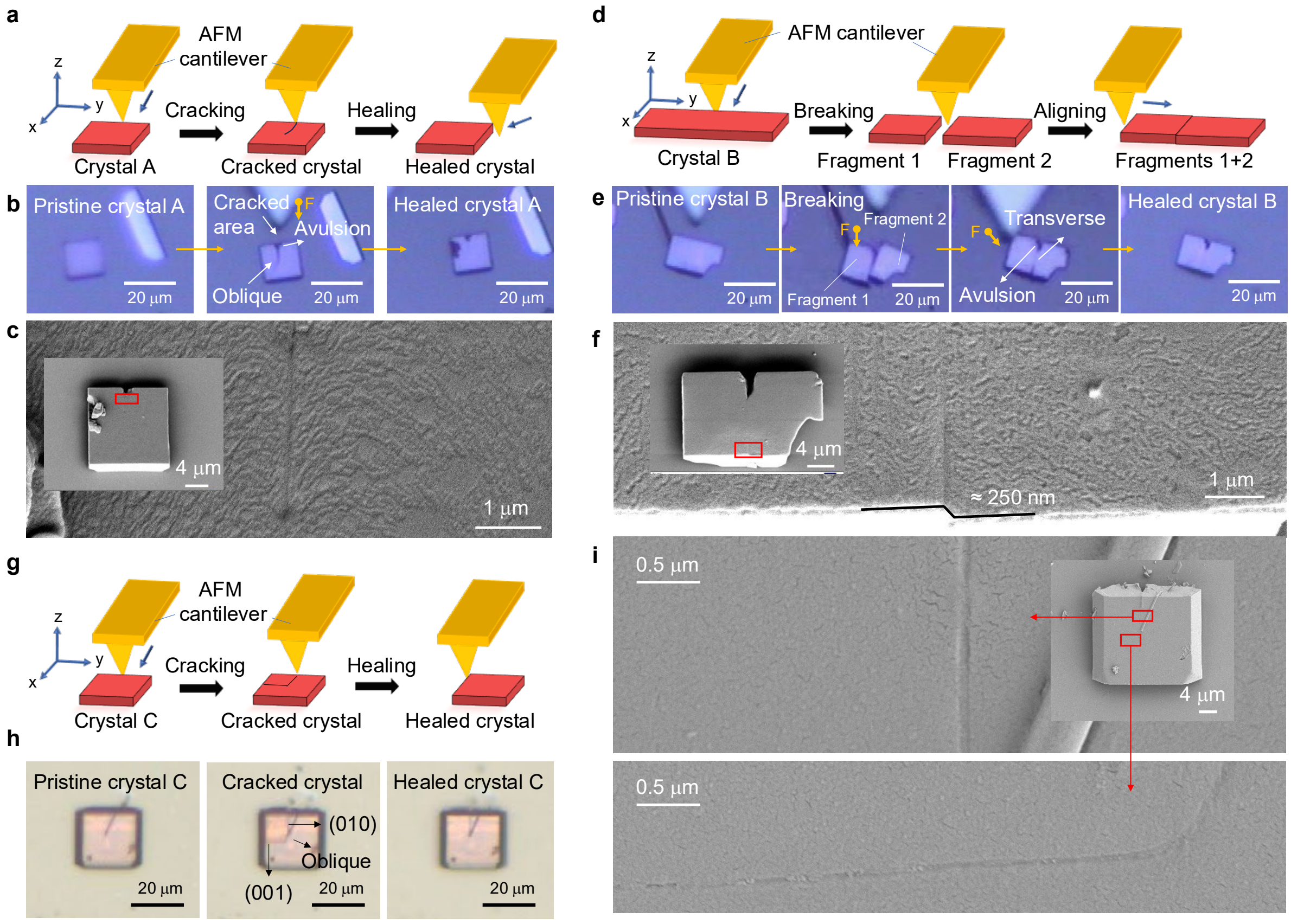

**Fig. 3 │ Self-healing of HDMAC microcrystals. a** Schematic illustration of self-healing in HDMAC microcrystal. **b** Sequential images of the mechanophotonics-assisted self-healing process. **c** FESEM image of the healed area at higher magnification. The inset shows the full crystal without a visible crack. **d** Schematic illustration of self-healed HDMAC microcrystal fragments. **e** Sequential images of the mechanophotonics-assisted self-healing process. **f** FESEM image of the healed area at higher magnification. The inset shows the entire crystal, with no visible crack at the healed area. **g** Schematic illustration of self-healing in two different planes of HDMAC microcrystal. **h** Sequential images of the mechanophotonics-assisted self-

healing process. **i** FESEM image of the healed area at higher magnification. The inset shows the entire crystal, with no visible crack at the healed area.

The healed microcrystal S2 was mechanically manipulated with the AFM tip in lateral (horizontal) and vertical directions. The two fragments moved together as a single body without separation, indicating restoration of mechanical continuity across the healed interface (Supplementary Fig. 8 and Supplementary video 3). To further establish the versatility of the self-healing process, a square-shaped HDMAC microcrystal C (19.3 × 19.3 $\mu m^2$) was selected to examine healing along distinct crystallographic planes (Fig. 3g). The crystal was deliberately fractured to generate oblique (non-displaced) cracks propagating along the (010) and (001) planes (Fig. 3h). The resulting fragments were then gently repositioned into their original orientation to facilitate healing. FESEM imaging revealed seamless healing of the oblique cracks along both the (010) and (001) planes, with only faint traces of the healed interfaces discernible at higher magnification (Fig. 3i). It was observed that nearly all the microcrystals successfully healed when the cracked surfaces remained clean and free of missing or/and added nanocrystalline debris (Supplementary Fig. 9 and Supplementary video 4,5). In contrast, when cleavage produced excessive nanocrystalline fragments at the fracture interface, effective healing was severely hindered.

The healing efficiency is governed by the degree of alignment and restoration of interfacial molecular contact between the fractured surfaces. Therefore, no single number can be given for this factor. Furthermore, the methods used to assess these two factors require improvement. As the 1D crystals exhibit optical waveguiding characteristics, we used optical loss as a parameter for the evaluation of healing efficiency (Fig. 4a,b). An elongated microcrystal D of length 271 μm was optically excited with the 405 nm continuous-wave laser at the position labeled 'L,' and the generated FL (Supplementary Fig. 10) self-propagated to the opposite terminal, and the spectrum was recorded at position 'R' (solid line, Fig. 4c). Later, a gentle force was applied using the cantilever tip on the (001) facet to introduce a transverse crack in the microcrystal. The microcrystal was carefully realigned, and a gentle force was applied to facilitate reattachment, resulting in a healed crystal (Fig. 4a,b). To assess the healing efficiency, the healed microcrystal was subjected to laser excitation at position 'L,' and the FL spectra were recorded as before. It is observed that there is no change in the FL intensity collected at position 'L' in both cases (before crack and after healing, Supplementary Fig. 10). However, the FL spectra collected at position 'R' showed a slight decrease in intensity after healing compared to the intensity before the crack (Fig. 4c). The percentage healing efficiency was calculated using the equation % healing efficiency $= \frac{s_1}{s_2} \times 100$, where $s_1$ and $s_2$ represent the areas under the guided FL spectra collected at position 'R' before cracking and after healing, respectively (Supplementary Fig. 10). The calculation indicated a healing efficiency of approximately 80%. To determine the average healing efficiency, experiments were conducted on six microcrystal optical waveguides with different dimensions. The healing efficiency ranged from 32% to 99%, with an average value of 88.5% (Supplementary Figs. S11–16). Variation in the measured healing efficiency arises due to light scattering at the cracked region, particularly at the region of missing crystal fragments.

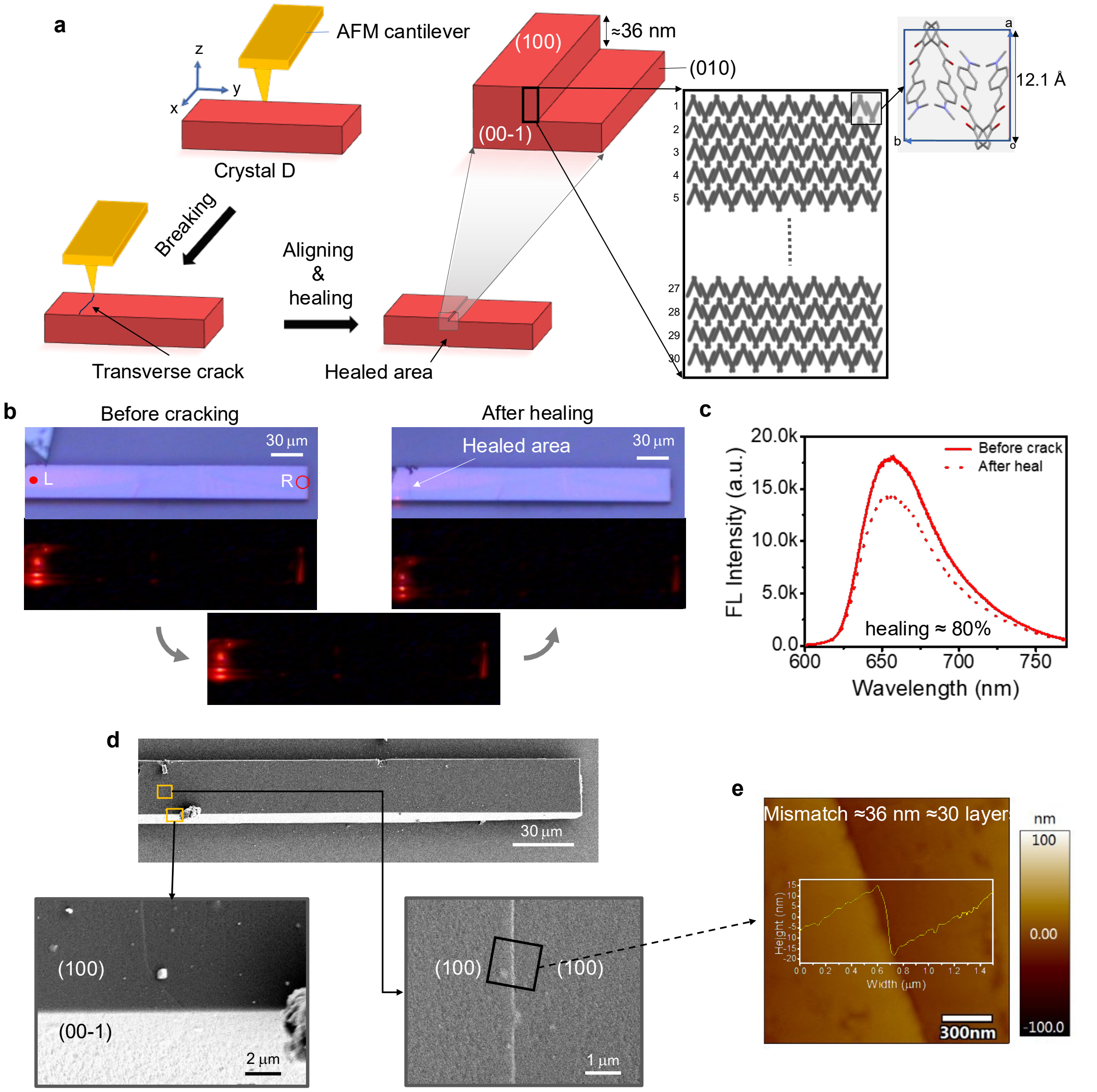

**Fig. 4 │ Self-healing optical waveguide. a** Graphical illustration demonstrating self-healing in HDMAC microcrystal. **b** Confocal and FL images of HDMAC microcrystal before and after healing. **c** Corresponding FL spectra with healing efficiency value. **d** FESEM images of the self-healed microcrystal with the magnified healed area. **e** AFM topographical height profile on the healed area.

Further, FESEM examination of the healed area at higher magnification showed a thin step defect. The step region points to a slight misalignment between the rejoined fragments in the (100) plane (Fig. 4d). The tapping mode AFM topographic image revealed a step height of ≈36 nm (inset, Fig. 4d). This height mismatch corresponds to a displacement of ≈30 unit cells in the crystallographic *b*-axis and ≈23,994 unit cells in the *c*-axis (Fig. 4a and Supplementary Note 2).

Interestingly, due to mirror-like light-reflecting facets, several microcrystal lasers[27] exhibited optical resonances in the FL spectra at the excitation point (Supplementary Fig. 17). The free spectral range (FSR) increased with decreasing crystal thickness, indicating a clear thickness dependence of the optical resonances (Supplementary Figs. 17,18), where the top and bottom crystal facets act as Fabry–Pérot (FP) type light-reflecting mirrors.

To investigate the self-healing capability of the photonic resonator, a nearly square-shaped HDMAC microcrystal E with dimensions of 31×28×1.8 $\mu m^3$ ($l$×$b$×$h$) was excited, and the crystal showed optical modes with a FSR of 17.8 nm (Fig. 5a,b,e,k). Subsequently, a longitudinal crack was intentionally created at the center of the microcrystal using an AFM cantilever tip, without completely separating the crystal into two pieces (Fig. 5c). Upon excitation at the cracked region, the FL spectrum exhibited a pronounced suppression of the optical resonances (Fig. 5f,k). The substantial disappearance of the resonance modes indicates a strong dependence of the cavity response on the crystal thickness and structural integrity of the resonator. Following healing, no discernible crack was visible in the optical image (Fig. 5d). When the microcrystal was excited at the healed region, the resulting FL spectrum exhibited recovery of optical resonances observed in the pristine crystal (Fig. 5g,k). The complete recovery of the resonance pattern demonstrates the remarkable self-healing capability of the HDMAC microcrystal resonator. To probe the structural integrity of the healed region beyond the surface, three-dimensional (3D) FL mapping was performed using a confocal optical microscope. Seven sequential *x*–*y* images were acquired along the *z*-direction, along the

(100) face, with an interlayer spacing of 257 nm (< λ/2 of FL) (Fig. 5h). The FL image of the layers

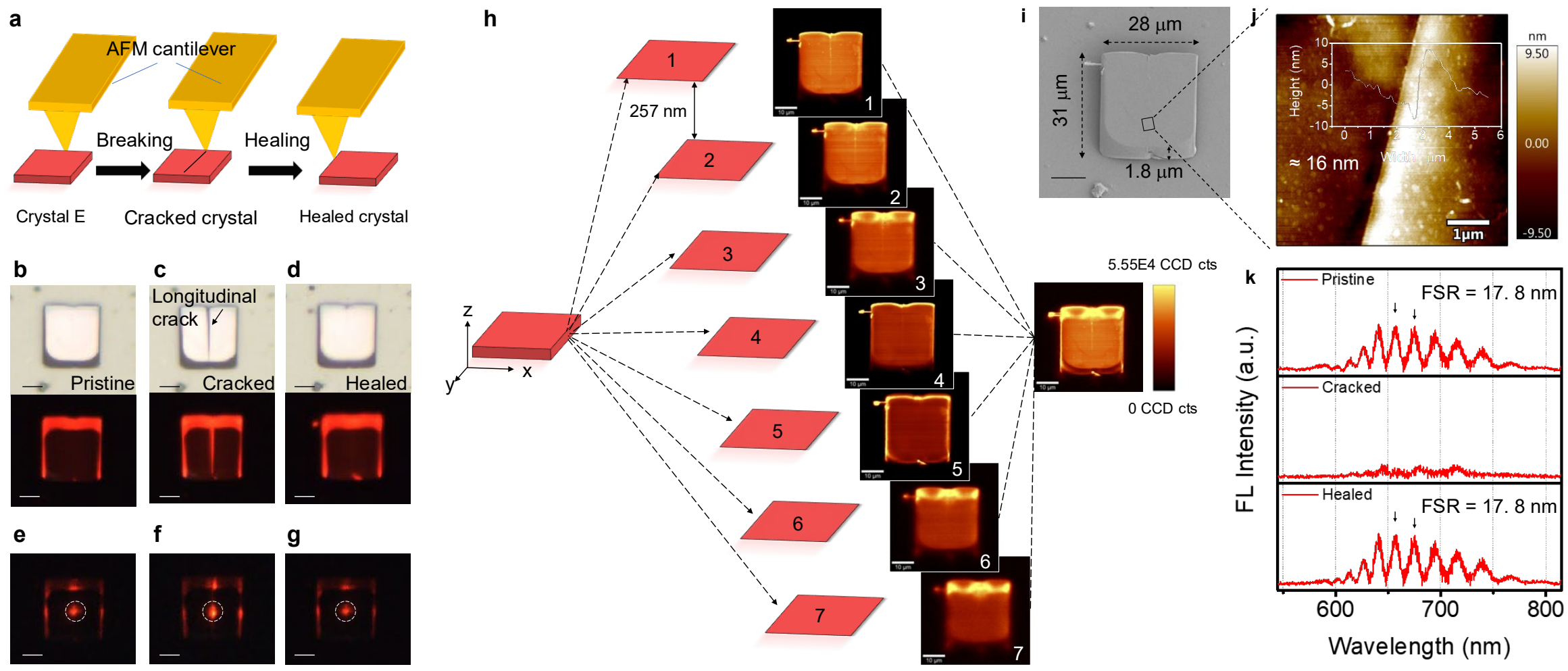

**Fig. 5 │ Self-healing optical resonator. a** Graphical illustration of self-healing in HDMAC microcrystal photonic resonator. **b-d** Confocal, FL images of HDMAC microcrystal optical resonator in the pristine, cracked, and healed states. **e-g** FL images showing the excitation position of the HDMAC microcrystal optical resonator in the pristine, cracked, and healed states. **h** Schematic illustration along with the confocal FL microscopic *z*-scan 3D mapping of healed microcrystal optical resonator. **i** FESEM image of the healed microcrystal optical resonator. All the scale bars are 10 μm. **j** AFM topographical height profile at the healed region. **k** FL background-subtracted resonance spectra recorded at the center of the pristine crystal resonator before cracking, the crack region, and the healed region.

provided the inner view of the healed region. Notably, a faint FL corresponding to the original crack was discernible only in the uppermost layers, whereas below ≈770 nm (see images 4-6 in Fig. 5h), substantial healing of the interior of the crystal was observed. The FESEM imaging at low magnification revealed a smooth and continuous surface after healing (Fig. 5i); however, AFM topography of the healed region revealed a ≈16 nm step defect at the healed region (Fig. 5j). The later data, accounts for the faint FL observed in the merged 3D FL image, likely arising from localized light scattering at the surface step (Fig. 5h). Notably, the healed crystal fully recovered its FP resonance characteristics, with the optical spectrum matching that of the pristine crystal (Fig. 5k), confirming the effective restoration of the optical functionality.

Investigating the local mechanical strength of the healed micro-region of the crystal is essential for evaluating the crystal's structural integrity following self-healing. However, research in this area remains limited. Therefore, we employed AFM-based nanoindentation to probe the mechanical response of the healed micro-region of the HDMAC crystal on its (100) surface (Fig. 6a). Nanoindentation measurements were performed on the pristine (A–C) and healed (D–G) regions using a conical diamond tip under an applied load of approximately 45 μN. At this load, the maximum indentation depths were 95 ± 1 nm and 115 ± 4 nm for the pristine and healed regions, respectively (Fig. 6b). The corresponding load (*P*)–displacement (*h*) curves exhibited distinct unloading profiles, indicating elastic recovery upon withdrawal of the indenter, alongside residual indentation after unloading (Fig. 6c). These responses demonstrate that both regions retain a measurable mechanical response under loading, with the healed region exhibiting greater penetration under the same applied load. The local mechanical properties extracted using the Oliver–Pharr method, yielded a Young's modulus (*E*) of 2.15 ± 0.13 GPa and hardness (*H*) of 0.175 ± 0.005 GPa for the pristine region, compared with $E$ = 1.49 ± 0.19 GPa and $H$ = 0.088 ± 0.047 GPa for the healed region. (Fig. 6d) The healed region exhibited a lower modulus, corresponding to approximately 66% retention of the pristine value. Taken together, the preserved surface morphology, measurable elastic recovery during unloading, and retention of substantial stiffness provide complementary evidence of mechanical integrity in the healed region. When considered alongside the recovery of optical functionality, these findings support the healing of the HDMAC crystal.

To gain molecular-level insight into the role of intermolecular interactions in the self-healing behavior of the HDMAC molecule, an energy framework analysis was conducted using the CrystalExplorer21 software.[30] The total intermolecular energy was decomposed into electrostatic, polarization, dispersion, and repulsion components, which were visualized as cylinders along the three crystallographic axes (Fig. 6e-g, and Supplementary Fig. 19). The graphical illustration of the self-healing planes and the Coulombic intermolecular energy framework presented in Fig. 6e,f reveals continuous attractive interaction pathways along the (010) and (001) planes (Fig. 2h). These asymmetric interactions exhibit energies of ≈ −36.4 and 45.2 kJ mol$^{-1}$ along the (010) plane and a stronger interaction of approximately −36.8 and 45.2 kJ mol$^{-1}$ along the (001) plane. The corresponding intermolecular contacts within the unit-cell packing, shown in Fig. 6g, demonstrate how each molecule is connected through an extended network of attractive interactions (C-H···O and interlayer dipole-dipole interactions[36]). Such intermolecular contacts can help the reapproach, molecular recognition, and reorganization of fractured crystal surfaces, thereby facilitating the self-healing of the crystal lattice. Overall, this analysis suggests that intermolecular interactions, particularly interlayer dipole-dipole interactions and C-H···O hydrogen bonding interactions, are one of the driving forces contributing to the self-healing property of the HDMAC molecule.

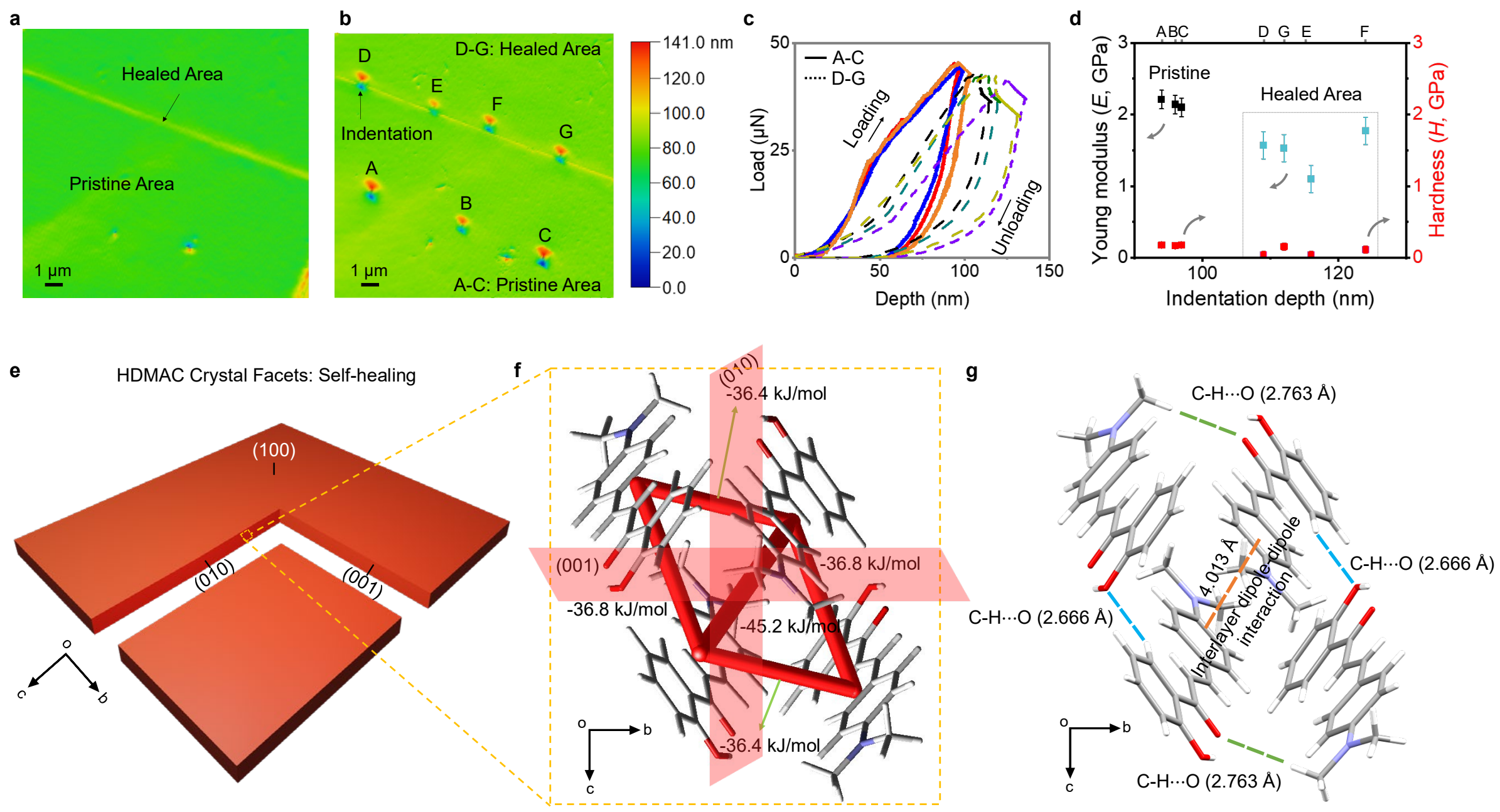

**Fig. 6 │ Nanoindentation studies of pristine and self-healed area and postulated healing mechanism.** AFM topographical images showing **a** (100) plane of the pristine and self-healed region prior to nanoindentation. **b** the residual indent within the pristine and self-healed area. **c** Representative load–displacement (*P*–*h*) nanoindentation curves recorded on the top plane for locations A–C, D–G. **d** Comparison of Young's modulus (*E*) and hardness (*H*) values between pristine and healed crystals. **e** Graphical illustration demonstrating the self-healing process along (010) and (001) planes. **f** The coulombic energy components of the intermolecular interaction energy visualized as energy frameworks, viewed along the crystallographic *a*-axis. Cylinders connecting molecular centroids represent pairwise interactions, with thickness proportional to interaction strength. Red cylinders represent attractive electrostatic interactions. The cylinder size and the cut-off energy are set to 150 and 10 kJ/mol, respectively. The plots were produced with CrystalExplorer21. **g** 3-D network unit cell packing structures with various intermolecular interactions for HDMAC crystals along the *a*-axis.

## Discussion

In summary, the microscale HDMAC single crystals show the ability to self-heal micromechanically-induced cracks (mechanophotonics) and completely fractured crystal segments while recovering their optical waveguiding and resonant properties. We propose various crack types as they are imperative for evaluating healing efficiency. The healing process was demonstrated on two accessible planes (001) and (010), highlighting the ability of these molecular crystals to repair damage under different crack types, including transverse, longitudinal, and oblique.

For practical self-healing nanophotonic devices, viz., waveguides, resonators, and lasers, the crystals should be on the order of a few microns, where even localized nanoscale damage can substantially affect optical performance. We used optical loss as a quantitative metric to evaluate healing efficiency, providing a practical approach to assess the functional recovery of photonic devices. Crystal dimensions, crack type, and propagation pathway, fragment realignment, and the underlying healing mechanism collectively determine the extent of recovery. Therefore, direct comparison of healing efficiencies across different molecular crystals with substantially different characteristics is not appropriate. Importantly, providing a single precise numerical value for diverse crack dynamics for the same crystal may give misleading healing efficiency.

In our study, healing was observed in microdomain crystals, whereas macro crystals (> 500 μm) did not exhibit healing under severe damage conditions, due to the presence of multiple crack pathways or their combinations. More importantly, multiple macroscopic crystal avulsions can disrupt fragment realignment and prevent effective healing across the fractured interface. Complementary FESEM and AFM characterization of the healed regions provides different morphological evidence of nanoscale cracks, emphasizing the importance of careful characterization. For 3D FL mapping, axial slicing at the nanoscale is useful for adequately assessing the structural integrity of the healed region. Furthermore, nanoindentation studies have traditionally not been performed on self-healed regions, leaving the mechanical properties of these regions poorly understood. The AFM-based nanoindentation measurements reveal that the healed regions exhibit approximately 50% greater penetration depth under the same applied load as pristine crystals, along with slight variations in elastic modulus and hardness. These findings highlight the importance of employing a combination of complementary techniques, including optical loss measurements, resonator characterization, surface and bulk microspectroscopy, and mechanical analysis of healed regions, to obtain a comprehensive and reliable assessment of healing efficiency. Overall, this work offers new opportunities for developing self-repairing and reconfigurable optical devices with enhanced reliability and operational lifetime.

## Methods

### Synthesis procedure of HDMAC

A solution of 2-hydroxyacetophenone (912 mg, 806.5 mL) in anhydrous ethanol (10 mL) was added under magnetic stirring to a prepared solution of sodium hydroxide (500 mg) in water (5 mL). Further, a solution of 4-dimethylaminobenzaldehyde (1 g) in anhydrous ethanol (12 mL) was added dropwise to the previous mixture. The solution was stirred at 25 °C for 24 hours, and then the precipitate was filtered to give a brown-orange solid. Finally, the solid was added to deionized water, filtered after standing overnight, and washed twice with deionized water. After drying, the coarse powder was recrystallized in ethanol.[27]

### Preparation of HDMAC microcrystals

For self-assembly, HDMAC (2 mg) was dissolved in Ethanol (1 mL) in a clean vial. The solution was heated rapidly until it became a clear orange color. Then, 50 µl of the solution was drop-cast onto a clean glass coverslip and left to evaporate slowly. Once the solvent evaporated, it resulted in 2D sheet/block-like microstructures.

### Solid-state absorbance and emission studies

The solid-state absorbance measurement was done on a Shimadzu UV-3600 spectrometer in a diffuse reflectance UV−visible (DR−UV−vis) mode. The reflectance spectra were converted to an absorbance using the Kubelka−Munk function. The solid-state emission spectrum was collected using a FLUOROMAX spectrofluorometer (HORIBA, Jobin Yvon).

### Single-crystal X-ray diffraction analysis

Single-crystal X-ray diffraction data were collected on a Rigaku Oxford XtaLAB ProPilatus3 R 200K-A detector system equipped with a CuKα (λ = 1.54184 Å) MicroMax-003 microfocus sealed tube operated at 50 kV and 0.6 mA. All data were collected at 298 K, and the data reduction was performed using CrysAlisPro software. The crystal structure was refined and solved using the OLEX software. Face indexing was done using CrysAlisPro software.

### Atomic force microscopy (AFM) studies

The surface topography analysis was performed on the NT-MDT Model Solver Pro M microscope using a class 2R laser at 650 nm with a maximum output of 1 mW. All calculations and image processing were carried out using the NOVA 1.0.26.1443 software provided by the manufacturer. The images were recorded in the tapping mode using a noncontact super sharp silicon cantilever (NSG 10_DLC) with a diamond-like carbon (DLC) tip purchased from NT-MDT, Moscow. The dimension of the tip is as follows: Cantilever length = 100 (±5) µm, Cantilever width 35 (±5) µm, and Cantilever thickness = 1.7-2.3 µm, Resonate frequency = 190-325 kHz, Force constant = 5.5-22.5 N/m, Chip size = 3.6×1.6×0.4 mm, Reflective side = Au, Tip height = 10-20 µm, DLC Tip curvature radius = 1-3 nm.

**Field-emission scanning electron microscopy (FESEM) studies**

The morphological analysis of the HDMAC microcrystals was performed using a Zeiss field-emission scanning electron microscope (FESEM) operating at 3-15 keV. All the experiments were performed after gold coating the samples.

**Nanoindentation studies using AFM**

Nanoindentation experiments were performed on a flat HDBP crystal surface using an Oxford Asylum AFM system. Indentation tests were carried out in indentation mode using a four-sided conical synthetic diamond probe (NM-TC, Adama Innovations) with a nominal half-angle of 45 ± 10°. The spring constant (k) of the AFM cantilever was determined using the empirical relation $k = AQf^{1.3}$, where the proportionality constant A ($8.296 \times 10^{-9}$) was provided by the manufacturer. The quality factor ($Q \approx 1100$) and resonance frequency ($f \approx 770$ kHz) were obtained from the system's auto-tuning routine.

The inverse optical lever sensitivity (InvOLS, 122 nm/V) was calibrated using the Sader method.[31,32] For this, the cantilever response was measured on a polished sapphire substrate, and the InvOLS was extracted by fitting the slope of the deflection displacement curve. Borosilicate glass was used as a reference material for mechanical calibration, with literature values of Young's modulus (82 GPa) and hardness (5.6 GPa). Young's modulus of the studied crystal was obtained using the Oliver–Pharr analysis implemented in Asylum Research software (v16.25.226), following the standard equations 1 and 2. Hardness values were calculated according to equations 3 and 4.[33,34]

**Nanoindentation calculations:**

i. Young modulus calculation using the Oliver-Pharr model

$$\frac{1}{E} = \frac{1-v^2}{\left(\frac{1}{E_r} - \frac{1-v_i^2}{E_i}\right)} \qquad \text{(Eq. 1)}$$

Where $E$ *and* $v$ are the Young's Modulus and Poisson's ratio of the material, $E_i$ and $v_i$ are Young's modulus and Poisson's ratio of the indenter tip. $E_r$ reduced modulus.

$$E_r = \frac{1}{2}\frac{\sqrt{\pi}}{\beta}\frac{S}{\sqrt{A_C}} \qquad \text{(Eq. 2)}$$

Young's modulus, where $S$ is stiffness dP/dh, $\beta$ is correction factor, $A_c$ = contact area.

ii. Hardness calculation

$$H = \frac{F}{A} \qquad \text{(Eq. 3)}$$

Where H is the hardness, F is the applied force in μN, and A is the area of the indent site.

$$A = \pi h^2 tan^2(\theta) \qquad \text{(Eq. 4)}$$

Where h is the depth of the indent, and θ is the half-angle of the cone-shaped diamond tip.

**Fluorescent (FL) 3D mapping studies**

Fluorescence 3D mapping was performed on a Wi-Tec alpha 300 AR laser confocal optical microscope (LCOM) equipped with a Peltier-cooled CCD detector (Model: DV401A-BVF-352; Serial: CCD-30423). A 532 nm laser source was used for excitation. A 100× objective was used for image collection. All measurements were performed under ambient conditions.

**Confocal microspectroscopy studies**

The HDMAC microcrystals were examined using a transmission mode setup of the Wi-Tec alpha 300 AR laser confocal optical microscope (LCOM) equipped with a Peltier-cooled CCD detector. Using a 300 grooves/mm blazed grating, BLZ = 750 nm, the system was configured with an accumulation time of 10 s, and the integration time was typically 0.5 s. Every spectrum is obtained from 10 averaged accumulations. A 405 nm diode laser was used for excitation. A 20× objective was used for spectra and image collection. All measurements were performed under ambient conditions.

**Hirshfeld Surface and Energy Framework Analyses**

**a The Hirshfeld Surface (HS) analyses:** The Hirshfeld Surface (HS) analyses were carried out employing the CrystalExplorer 21 program.[30,35] The $d_{norm}$ surface was mapped with the color scale in the range −0.02 au (red, au = atomic unit) to 1.4 au (blue). The $d_{norm}$ property is a symmetric function of distances to the surface between nuclei inside and outside the Hirshfeld surface ($d_i$ and $d_e$, respectively), relative to their respective van der Waals radii. The regions with red and blue colors on the dnorm represent the shorter and longer inter contacts, while the white color indicates the contacts around the van der Waals radii.

**b Energy frameworks analysis:** The energy frameworks in HDMAC were constructed by using Crystal Explorer21, and interaction energies were calculated using the B3LYP hybrid functional with the DGDZVP basis set, where semiempirical dispersion was included using D3 version of Grimme's dispersion.[30] The total intermolecular interaction energy ($E_{tot}$) is the sum of four energy terms: electrostatic ($E_{ele}$), polarization ($E_{pol}$), dispersion ($E_{disp}$), and repulsion ($E_{rep}$). The relative strengths in interaction energies in individual directions are represented by cylinder-shaped energy frameworks. Insignificant contacts with energies below the threshold of 10 kcal/mol have been omitted from the original figure for clarity.

**Data availability**

All data are available from the authors on request. The authors declare that all other data supporting the findings of this study are provided in the Supplementary Information.

**Acknowledgements**

R.C. thanks JC Bose Grant (ANRF/JBG/2026/000116/CS and DST/INT/RUS/RSF/P-71/2023(G) for funding. M.C. thanks CSIR, K.Y. thanks UGC, and A.K. thanks PMRF for the fellowship.

**Author information**

Authors and Affiliations

**School of Chemistry and Centre for Nanotechnology, University of Hyderabad, Gachibowli, Hyderabad 500046, Telangana, India.**

Melchi Chosenyah, Vuppu Vinay Pradeep, Kapil Yadav, Ankur Khapre, Biswajit Kumar Barman, Rajadurai Chandrasekar

**Department of Chemistry, School of Chemical Sciences, Central University of Karnataka, Kadaganchi, Aland Road, Kalaburagi - 585367, Karnataka, India.**

Raghavaiah Pallepogu

Contributions

V.V.P. synthesized, carried out photophysical studies, collected the SCXRD data of the macrocrystal, and carried out initial self-healing experiments. M.C. carried out all experimental work, collected data, analyzed, and prepared the preliminary manuscript under the supervision of R.C. M.C. performed the FL 3D mapping studies with the help of K.Y. A.K. carried out nanoindentation studies using AFM. B.K.B. performed Hirshfeld surface and energy framework analysis. R.P. collected single-crystal XRD data of the microcrystal. R.C. supervised the work, provided guidance on research direction, planned the experiments, interpreted the results, provided critical suggestions, and prepared the final manuscript. All authors discussed the results and approved the final version of the manuscript.

Corresponding author

Correspondence to Rajadurai Chandrasekar

## Ethics declarations

Competing interests

The authors declare no competing interests.

## Supplementary information